# K-band LiNbO$_3$ A3 Lamb-wave Resonators with Sub-wavelength Through-holes

Shu-Mao Wu, Hao Yan, Chen-Bei Hao, Zhen-Hui Qin, Si-Yuan Yu, Yan-Feng Chen

*Abstract*— Addressing critical challenges in Lamb wave resonators, this paper presents the first validation of resonators incorporating sub-wavelength through-holes. Using the A3 mode resonator based on a LiNbO$_3$ single-crystal thin film and operating in the K band as a prominent example, we demonstrate the advantages of the through-hole design. In the absence of additional processing steps, and while maintaining device performance—including operating frequency, electromechanical coupling coefficient, and quality factor—without introducing extra spurious modes, this approach effectively reduces the ineffective suspension area of the piezoelectric LN film, potentially enhancing mechanical and thermal stability. It also standardizes etching distances (and times) across various Lamb wave resonators on a single wafer, facilitating the development of Lamb wave filters. The versatility of the through-hole technique, with relaxed constraints on hole geometry and arrangement, further highlights its significance. Together with the other advantages, these features underscore the transformative potential of through-holes in advancing the practical implementation of Lamb wave resonators and filters.

## I. INTRODUCTION

With the rapid advancement and deployment of satellite technologies, communication systems operating in frequency bands such as the Ku-band (12-18 GHz), K-band (18-26.5 GHz), and Ka-band (26.5-40 GHz) are facing increasingly stringent requirements [1]. Filters in these bands must not only function effectively at higher frequencies but also satisfy demands for compact size, wide bandwidth, low loss, and high stability. Existing filtering solutions, including LTCC (Low-Temperature Co-fired Ceramic), IPD (Integrated Passive Device), and traditional bandpass electromagnetic filters, face significant challenges in meeting these criteria. [2-4].

In recent years, Lamb wave acoustic resonators based on LiNbO$_3$ (LN) single-crystal thin films have emerged as a promising solution to the challenges of developing high-frequency, high-performance filters. The exceptional piezoelectric properties of LN, combined with the high-frequency characteristics of Lamb waves and the advancements in single-crystal thin-film technology, enable LN Lamb wave resonators to effectively balance high-frequency, large electromechanical coupling coefficients ($k_t^2$), and miniaturization [5]. Among LN Lamb wave resonators, the first-order asymmetric mode (A1) resonators exhibit the highest electromechanical coupling coefficient, exceeding 40%. Although higher-order asymmetric modes (A3, A5, A7, etc.) have lower coupling coefficients, they allow for higher frequency filtering in LN thin films of the same thickness, reaching frequencies of tens GHz or even higher [6].

The development of LN A-mode resonators and filters primarily focuses on enhancing key performance parameters, including the electromechanical coupling coefficient (which affects bandwidth), operating frequency, and quality factor (Q) [7]. Yang et al. devised and fabricated Z-cut LN A1 resonators with $k_t^2$ exceeding 26% and developed subsequent A1 filters at near 4.5 GHz with a bandwidth of 10% [8][9]. Additionally, Wu et al. developed a 6.2 GHz A1 filter with over 10% bandwidth on a single wafer using Z-cut LN thin films of varying thicknesses [10]. Meanwhile, Plessky et al. developed various A1 resonators with frequencies covering 3-7 GHz on Z-cut LN thin films, *i.e.*, the XBARs, with $k_t^2$ exceeding 25% [11], [12]. Building on XBAR technology, Resonant developed ultra-wide bandwidth filters for various frequency bands, including n77, n79, and Wi-Fi 6E [13], [14]. Furthermore, Lu et al. utilized 128° Y-cut LN thin films to elevate the $k_t^2$ of A1 resonators to 46%, laying the groundwork for filters with larger bandwidth [15]. Yang et al. employed Y-cut LN thin films to develop A1 resonators with Q exceeding 3000, although operated at a relatively low frequency of ~1.7 GHz [16]. In contrast, Gu et al. successfully increased the Q of the XBARs to over 1000 at higher frequencies of ~4.7 GHz by incorporating reflection gratings [17]. LN A resonators are also developing towards higher frequencies. By utilizing higher-order modes, researchers have achieved A3 filters in the X band at 8.4 GHz [18], A3-mode resonators with a resonant frequency of 20.3 GHz [19], A5 filters in the Ku band at 14.7 GHz, and A7 filters in the K band at 19 GHz [20]. The frequency can be increased by continuously reducing the thickness of the LN thin films. Yang et al. reported LN A1 resonators operating at ~25 GHz with electromechanical coupling coefficient of 16% [21]. Barrera et al. developed A1 filters of 23.5 GHz and bandwidth up to 18.2% [22]. Xie et al. created sub-THz resonators approaching 100 GHz and preliminarily designed a filter based on such resonators [23].

Although LN Lamb wave resonators offer significant advantages and are advancing rapidly, they still face substantial challenges related to mechanical stability, temperature stability, power capacity, and preparation costs due to their specific suspension structures [24]. Currently, there are two primary process routes for fabricating LN Lamb resonators and filters, with the main distinction being in the suspension method of the LN thin films. The first route involves removing the underlying Si (i.e., the substrate) or $SiO_2$ (i.e., the box layers) from beneath the LN thin film. This is achieved through release windows on the surface of the LN thin film using BOE etchant, HF, or $XeF_2$ gases [8],[19]. The second route eliminates the need for release windows by directly removing the substrate (typically Si) with a thickness of several hundred microns from the back side of the resonator substrate using a deep silicon etching process [12],[25]. Both routes have their limitations. The primary issue with the first route is that the etching processes using BOE solution, HF, or $XeF_2$ gases are isotropic, which can lead to ineffective suspension areas around the perimeter of the resonators. More importantly, resonators of different sizes require varying etching times, complicating the processing of multiple resonators on the same wafer, such as when fabricating filters. The second route, while avoiding the issue of isotropic etching, involves additional steps for back etching. This increases costs and suffers from lower accuracy in back alignment lithography.

This paper aims to address the key challenges faced by existing LN Lamb wave resonators. By introducing subwavelength through-holes on the surface of the LN thin film, the limitations of the first route are effectively resolved: (1) Ineffective suspension areas of the LN thin films are minimized without requiring additional processing steps, improving the resonators' mechanical stability and heat dissipation; (2) Etching times for resonators of different sizes are standardized, facilitating the fabrication of filters; (3) The resonator performance remains unaffected, and may even be enhanced, including the electromechanical coupling coefficient, frequency, quality factor, without introducing spurious modes. In this paper, we take the A3 resonators prepared on Z-cut LN thin film as an example. Compared with the resonators without through-holes, the resonators with through-holes have the same operating frequency (over 21GHz), electromechanical coupling coefficient (over 4%), and no additional spurious modes. We also demonstrated the wide applicability and design flexibility of through-holes. This method, which provides comprehensive benefits to Lamb wave resonators throughout their preparation and performance, is anticipated to significantly enhance the practical application of both Lamb wave resonators and filters.

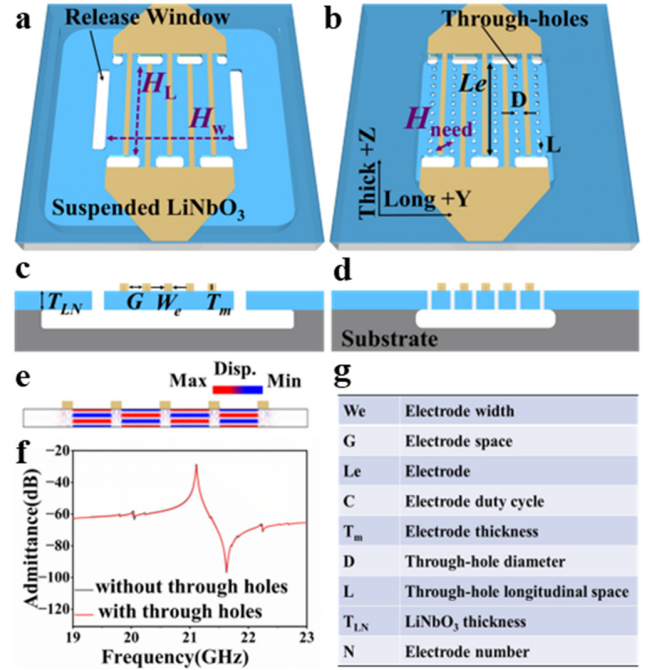

**Fig. 1. A3 resonator in a suspended LiNbO₃ thin film.** (a) Top view without through-holes, (b) Top view with through-holes, (c) cross-sectional view without through-holes, (d) cross-sectional view with through-holes, (e) displacement distributions at A3 resonances, (f) Simulated admittance spectra of Design I-1 & I-2 and (g) annotation of key parameters.

## II. ADVANTAGES OF THROUGH-HOLES FOR LAMB WAVE RESONATORS & EXEMPLARY EXPERIMENTAL DESIGN

Figs. 1(a) and 1(c) present the top view and cross-sectional schematic diagrams of a conventional A3 resonator without through-holes, while Figs. 1(b) and 1(d) illustrate the top view and cross-sectional schematic diagrams of the resonator featuring through-holes. The figures show that identical round through-holes are uniformly distributed across the surface of the LN thin film and between the interdigitated electrodes. The diameter of each through-hole is denoted as *D*, with a longitudinal spacing $L$ between adjacent through-holes. The interdigital electrodes are oriented perpendicular to the Y direction of LN, with a width $W_e$, an aperture length $L_e$, and a spacing $G$. The electrode duty cycle is $C$ (where $C = W_e / (W_e + G)$), and the total number of electrodes is $N$. The thickness of the Z-cut LN thin film is 260 nm, while the electrodes are made of Au with a thickness of 50 nm.

When preparing the resonator using release windows and isotropic etching (the first route mentioned above), the minimum etching distance required to fully suspend the LN thin film is approximately equal to the distance between the two outer release windows. This distance is typically represented by the horizontal spacing $H_W$ or the vertical spacing $H_L$, as indicated by the purple dotted line in Fig. 1(a), depending on the resonator design.

During the resonator fabrication process, it is ideal to arrange resonator with the same or similar needed etch distance ($H_{need}$) on one wafer for consistent processing. However, this is not always feasible in practice. For instance, in filter design, multiple resonators with a wide range of different $C_0$ are needed to optimize filter performance. This necessitates adjusting the structural parameters of every resonator, including the $N$, $Le$, $G$, and $T_{LN}$. Common approaches to achieve this include: 1) combining multiple smaller resonators in parallel to form a larger resonator unit and adjusting $C_0$ by the number of parallel units; 2) varying $N$ of the interdigital electrodes; and 3), adjusting $Le$ of the interdigital electrodes. To build filters with the best performance, multiple parameters are often adjusted simultaneously, leading to significant variations in $H_{need}$ of different resonators. If the $H_{need}$ for each resonator in the filter varies significantly, they must be processed according to the maximum $H_{need}$. This results in a considerable amount of ineffective LN suspension around the relatively small resonators, which can adversely affect the device's stability—both mechanical and thermal—as well as its compactness.

The through-hole structure effectively addresses the issue of varying $H_{need}$. As depicted in Fig. 1(b), the $H_{need}$ for a resonator with through-holes is determined by the diagonal length of the quadrilateral formed by four adjacent through-holes, as indicated by the purple solid line in the figure. Compared with resonators without through-holes, $H_{need}$ is significantly reduced, which minimizes or even eliminates the ineffective suspended area around the resonator. More importantly, the $H_{need}$ can be controlled by the shape, size, spacing, and other parameters of the through-holes. This allows for a consistent $H_{need}$ across different resonators on the same wafer, greatly facilitates their design and fabrication.

TABLE I
DESIGN PARAMETERS OF THE A3 RESONATORS
(DESIGN I-1 TO VIII-2)

| Parameter | Through holes | We | Le | G | N | T |
|---|---|---|---|---|---|---|
| Design I-1 | // | 2μm | 100μm | 8μm | 5 | // |
| Design I-2 | Yes | 2μm | 100μm | 8μm | 5 | // |
| Design II-1 | // | 2μm | 200μm | 8μm | 5 | // |
| Design II-2 | Yes | 2μm | 200μm | 8μm | 5 | // |
| Design III-1 | // | 2μm | 300μm | 8μm | 5 | // |
| Design III-2 | Yes | 2μm | 300μm | 8μm | 5 | // |
| Design IV-1 | // | 2μm | 100μm | 8μm | 25 | 5 |
| Design IV-2 | Yes | 2μm | 100μm | 8μm | 25 | 5 |
| Design V-1 | // | 2μm | 100μm | 8μm | 25 | // |
| Design V-2 | Yes | 2μm | 100μm | 8μm | 25 | // |
| Design VI-1 | // | 4μm | 100μm | 6μm | 25 | 5 |
| Design VI-2 | Yes | 4μm | 100μm | 6μm | 25 | 5 |
| Design VII-1 | // | 2μm | 200μm | 8μm | 25 | 5 |
| Design VII-2 | Yes | 2μm | 200μm | 8μm | 25 | 5 |
| Design VIII-1 | // | 2μm | 300μm | 8μm | 35 | 7 |
| Design VIII-2 | Yes | 2μm | 300μm | 8μm | 35 | 7 |

To thoroughly verify the advantages of through-hole resonators, we designed a range of A3 resonators with varying parameters. The specific parameters for these resonators are detailed in Tables I. In Design I-1 to VIII-2, the through-holes are uniformly distributed at the center of adjacent Au electrodes, are round with a diameter $D$ of 1 μm, and have a longitudinal spacing $L$ of 10 μm.

The commercial multiphysics simulation software COMSOL was used to model the full 3D structures of the Design I-1 and I-2 resonators, as shown in Fig. 1(f). The black curve represents the admittance spectrum of the conventional resonator without through-holes, while the red curve shows the spectrum for the resonator with through-holes. The Fig. 1(e) displays the resonator's displacement distribution. The simulation results indicate that the spectra are nearly identical within the shown frequency range, providing a preliminary indication that the through-hole structure may not adversely affect the resonator's performance.

III. PREPARATION OF A3 RESONATORS: WITH AND WITHOUT THROUGH-HOLES

To validate our design, we fabricated the resonators according to the specified process flow, as depicted in Fig. 2. The 260 nm Z-cut LN single-crystal thin film was sourced from the commercial supplier NANO LN.

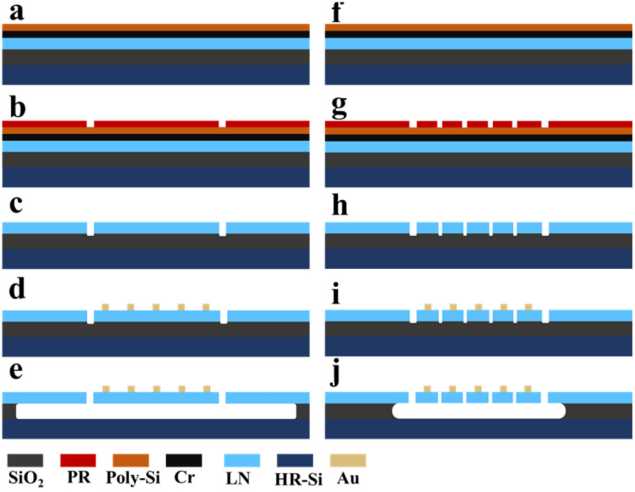

**Fig. 2.** Fabrication process of the LiNbO$_3$ A3 resonators (a-e) without through-holes, (f-j) with through-holes

Fig. 2(a)-(e) illustrate the preparation process for resonators without through-holes, while Fig. 2(f)-(j) depict the process for resonators with through-hole structures. Initially, a 320 nm thick chromium (Cr) layer is deposited by electron beam evaporation to serve as a hard mask for etching the LN film. Following this, a 500 nm thick silicon dioxide (SiO$_2$) layer is deposited via plasma-enhanced chemical vapor deposition (PECVD) to act as an etching mask for the Cr layer. Photolithography is then used to define the release windows and through-holes in the resonator. The LN film is subsequently etched using inductively coupled plasma reactive ion etching (ICP-RIE) to form the release windows and through-holes. Next, a 50 nm thick gold (Au) electrode is deposited by electron beam evaporation and patterned using a lift-off process. Finally, the SiO$_2$ layer at the bottom of the LN film is removed with a buffered oxide etch (BOE) solution, and the wafers are dried using a critical point dryer.

Overall, the process flow for fabricating resonators, whether they include through-holes or not, remains consistent, as the addition of through-hole structures does not necessitate any additional steps. There are two differences in the processes: First, the etching of the LN film. As shown in Fig. 2(c) and Fig. 2(h), resonators with through-holes require additional etching to create these through-holes. However, this additional etching can be performed simultaneously with the etching of the original release window, without adding extra steps. The second difference concerns the release of the LN film. As illustrated in Fig. 2(e) and Fig. 2(j), the through-holes provide extra channels for the etching process, which shortens the etching distance ($H_{need}$) and reduces the ineffective suspension area of the LN around the resonators.

We characterized the morphology and line width of the fabricated devices. Fig. 3 presents SEM images of the resonators for Design I-1 and I-2. Fig. 3(a) and 3(b) depict the core structural areas of the resonators with and without through-holes, respectively. The images demonstrate that the resonator surfaces are clean and well-defined, with no apparent defects associated with the through-hole structures. The interdigital electrodes exhibit a uniformly dense and high-quality surface, with line widths closely matching the design specifications. The contours of the etched release windows and through-holes are clearly defined, and the through-holes are periodically distributed across the LN surface and centrally within the interdigital electrodes, with their characteristic line widths conforming to the design.

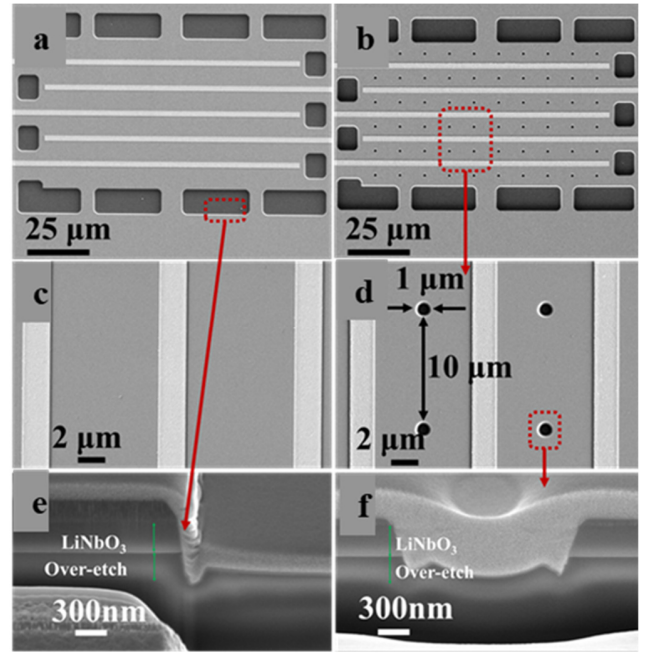

**Fig. 3.** SEM images of the fabricated A3 resonators for Design I-1 and I-2. (a) Core area of the resonator without through-holes, (b) Core area of the resonator with through-holes, (c) Zoomed-in view of the resonator's electrodes without through-holes, (d) Zoomed-in view of the resonator's electrodes with through-holes, (e) Side view of the release window, and (f) Side view of the through-holes.

The suspended area of the resonator is visible through its "undercut". Fig. 4 presents optical microscope images of Design V-1 and Design V-2 after the release process. For resonators without through-hole structures, the BOE solution can only etch SiO$_2$ from the release windows outside the resonator, progressing from the edges toward the center. This extended etching path results in a large ineffective suspension area, as shown in Fig. 4(a). Conversely, for resonators with through-hole structures, the BOE solution can etch SiO$_2$ simultaneously at all locations, including the center and edges of the resonator, through the through-holes. This significantly shortens the etching distance, reduces the ineffective suspension area, as depicted in Fig. 4(b). The resonators shown in Fig. 4(b) and Fig. 4(c) were processed simultaneously on the same wafer. The images reveal that, after the same duration of BOE processing, the resonator with through-holes is fully suspended, whereas the central area of the resonator without through-holes remains largely unsuspended, as highlighted by the red dotted box in Fig. 4(c).

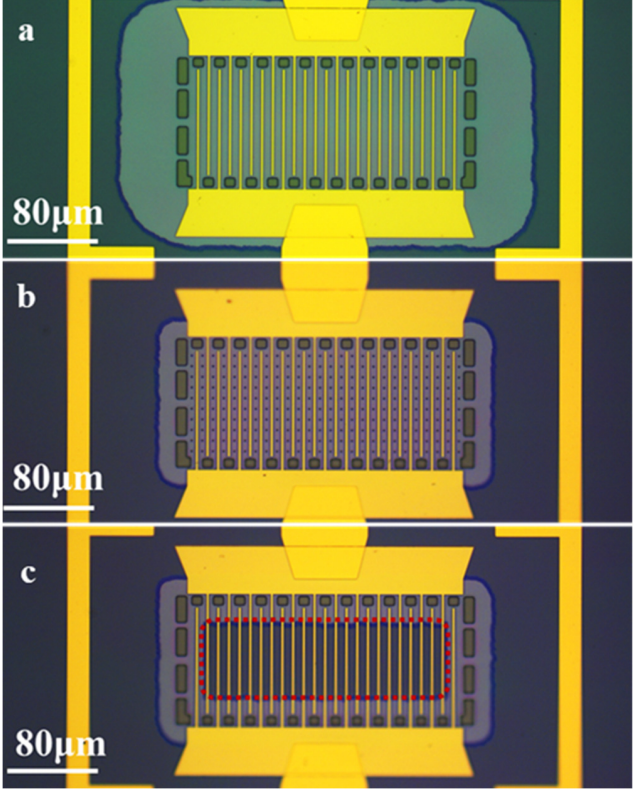

**Fig. 4.** Optical microscope images of the fabricated resonators for Design V-1 (without through holes) and Design V-2 (with through holes). (a) Design V-1: fully suspended, (b) Design V-2: fully suspended, and (c) Design V-1: not fully suspended.

## IV. PERFORMANCE OF A3 RESONATORS: WITH AND WITHOUT THROUGH HOLE

Admittance spectra of all fabricated A3 mode resonators were measured using an Agilent M9375A PNA vector network analyzer with an MPI GSG200 RF probe. To ensure accuracy, meticulous short-open-load-through (SOLT) calibration and de-embedding procedures were implemented during the measurement process.

In designing acoustic filters, achieving optimal filter performance requires impedance matching, which necessitates a resonator with a specific value of $C_0$. One effective method to control $C_0$ is by adjusting the aperture of the interdigital electrodes [12],[18]. We first assessed the effect of varying the aperture size of the interdigital electrode (Designs I-1-III-2) on the resonator's performance. The results are shown in Fig. 5, which includes optical microscope images of each device after suspension. Key performance parameters of the resonators are summarized in Table II. The $k_t^2$ is calculated by the formula $k_t^2=\pi^2/4*((f_p^2-f_s^2)/f_p^2)$ and the quality factor is expressed by $Q_{3db}$ ($Q_{3db}=\Delta f_{3db}/f_s$).

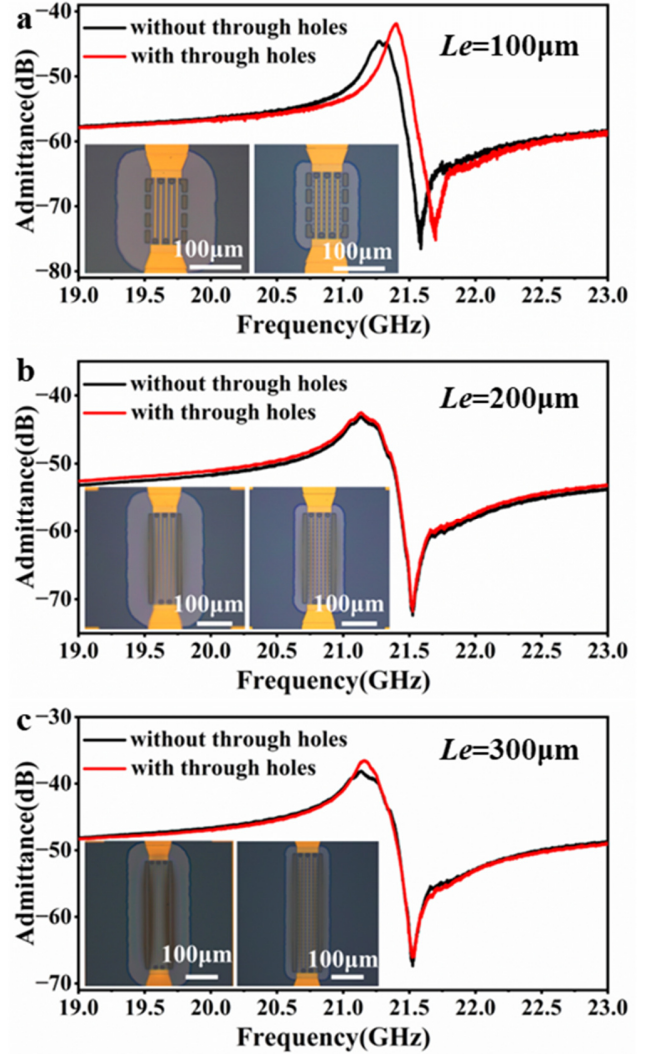

**Fig. 5.** Measured admittance spectra of the A3 resonators: (a) Design I-1 & I-2, (b) Design II-1 & II-2, (c) Design III-1 & III-2. Insets are optical microscope images of the corresponding resonators without (left) and with (right) through-holes. The parameters of the resonators are detailed in Table I.

As illustrated in Fig. 5, when Le=100 μm, there is a noticeable frequency deviation between the resonator with through-holes and the one without. However, several other key performance parameters remain unaffected. This frequency deviation is primarily due to the uneven thickness of the LN film. At Le=200μm and Le=300μm, the admittance spectra of the resonators with and without through-holes nearly overlap, indicating that the through-hole structure has little impact on the key parameters of the resonator. The data presented in Fig. 5 and Table II indicate that when modifying the aperture size, the through-hole structure effectively reduces the ineffective suspension area of the resonator. Specifically, the overall suspension areas of the resonators are reduced by approximately 50% to 60%, which inevitably enhances their stability.

TABLE II
EXTRACTED KEY PARAMETERS FROM DEVICE MEASUREMENT
(DESIGN I-1-III-2)

| Device | $f_s$(GHz) | $k_t^2$ | $Q_s$ | $Q_p$ | $H_{need}$(μm) | Size(μm) |
|---|---|---|---|---|---|---|
| Design I-1 | 21.26 | 3.77% | 107 | 770 | 58 | 190*230 |
| **Design I-2** | **21.40** | **3.97%** | **189** | **867** | **15** | **120*160** |
| Design II-1 | 21.11 | 4.59% | 68 | 467 | 58 | 190*330 |
| **Design II-2** | **21.11** | **4.59%** | **68** | **467** | **15** | **120*260** |
| Design III-1 | 21.14 | 4.44% | 66 | 497 | 58 | 190*430 |
| **Design III-2** | **21.13** | **4.33%** | **130** | **460** | **15** | **120*360** |

TABLE III
EXTRACTED KEY PARAMETERS FROM DEVICE MEASUREMENT
(DESIGN VI-1-VIII-2)

| Device | $f_s$(GHz) | $k_t^2$ | $Q_s$ | $Q_p$ | $H_{need}$(μm) | Size(μm) |
|---|---|---|---|---|---|---|
| Design VI-1 | 21.43 | 3.74% | 133 | 232 | 56 | 455*230 |
| **Design VI-2** | **21.43** | **3.94%** | **133** | **232** | **15** | **685*160** |
| Design VII-1 | 21.24 | 4.01% | 68 | 240 | 58 | 460*430 |
| **Design VII-2** | **21.26** | **3.92%** | **105** | **342** | **15** | **395*360** |
| Design VIII-1 | 21.25 | 4.44% | 76 | 254 | 58 | 595*430 |
| **Design VIII-2** | **21.29** | **3.98%** | **201** | **257** | **15** | **525*360** |

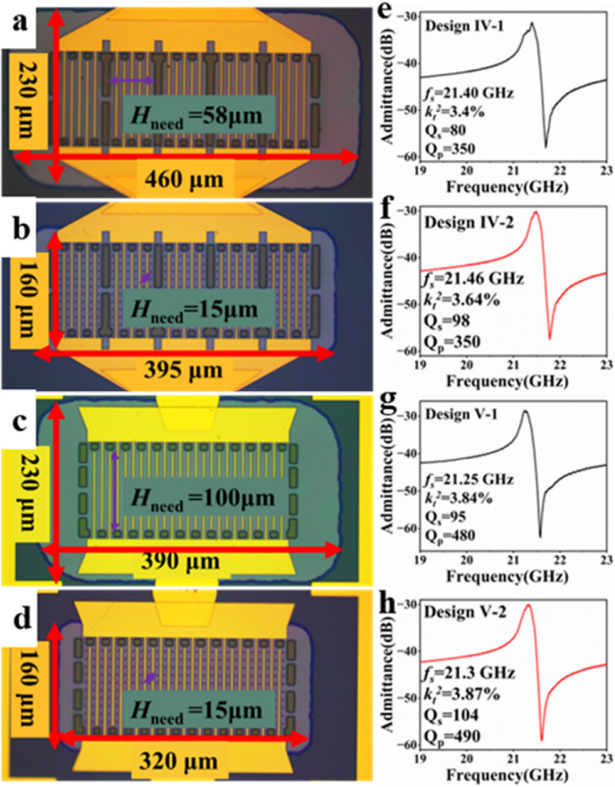

**Fig. 6.** Optical microscope images of A3 resonators: (a) Design IV-1, (b) Design IV-2, (c) Design V-1, (d) Design V-2. Measured admittance spectra: (e) Design IV-1, (f) Design IV-2, (g) Design V-1, (h) Design V-2. The parameters of these resonators are detailed in Table I.

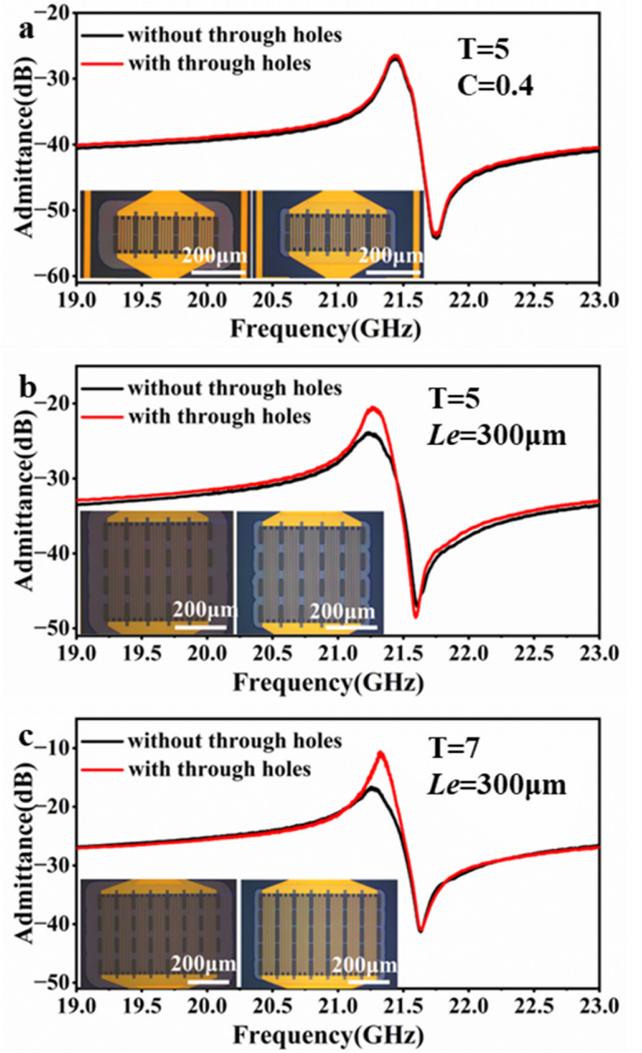

**Fig. 7.** Measured admittance spectra of the A3 resonators: (a) Design VI-1 & VI-2, (b) Design VII-1 & VII-2, (c) Design VIII-1 & VIII-2. Insets are optical microscope images of the corresponding resonators without (left) and with (right) through-holes. The parameters of the resonators are detailed in Table I.

Fig. 6 displays optical microscope images and measured results for the resonators of Design IV-1, Design IV-2, Design V-1 and Design V-2, with key design parameters detailed in Table I. Comparing (a) with (b) and (c) with (d), the resonators with through-holes show a reduced etch distance, leading to a 40% decrease in ineffective suspension area. Fig. 6(a) and (c) show that resonators without through-holes have different etch distances $H_{need}$: 58 μm for Design IV-1 and 100 μm for Design V-1. The same wafer processing complicates Design IV-1's processing and increases its ineffective suspension area. In contrast, through-holes in Designs IV-2 and V-2 standardize etch distances, simplifying the release process, as shown in Fig. 6(b) and (d). Fig. 6(e)–(h) shows that the admittance spectra of the four resonators are similar and that the through-hole structure does not add spurious modes. Performance parameters shown in the Fig. 6 indicate that the resonators' key performance remains unaffected.

We further investigated parallel-designed resonators by varying the duty cycle, number of parallel units, and aperture size, with and without through-holes. Design parameters are detailed in Table I, and results are shown in Fig. 7. Insets provide optical microscope images, and Table III summarizes key performance metrics. The admittance spectra for resonators with and without through-holes largely overlap, with no additional spurious modes introduced. The through-hole structure does not negatively impact performance, and it reduces the overall suspension area by approximately 30%, as indicated in Table III.

## V. PERFORMANCE OF A3 RESONATORS WITH DIFFERENT THROUGH HOLES

Using resonator V-2 as a basis, we investigated the effects of through-hole shape, position, spacing, and diameter on performance. Admittance spectra for these variations are shown in Figs. 8, 9, and 10, which also include optical microscope and magnified SEM images of the resonators. Key performance parameters are summarized in Table IV.

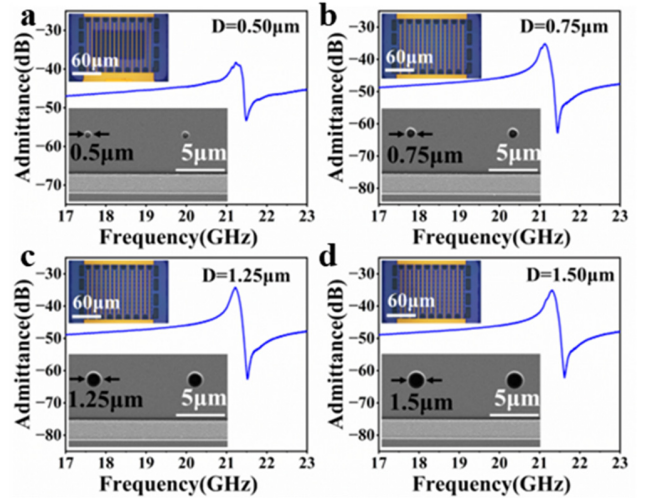

**Fig. 9.** Measured admittance spectra of the A3 resonators with different through-hole diameters (a) D=0.50μm, (b) D=0.75μm, (c) D=1.25μm, (d) D=1.50μm. Insets are (top) optical microscope images of the corresponding resonators and (bottom) zoomed-in SEM images.

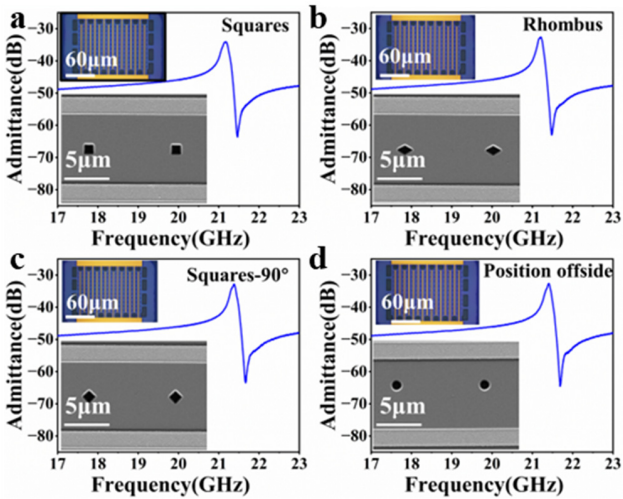

**Fig. 8.** Measured admittance spectra of the A3 resonators with different through-hole shapes (a) Squares, (b) Rhombus, (c) Squares-90, (d) position offside. Insets are (top) optical microscope images of the corresponding resonators and (bottom) zoomed-in SEM images.

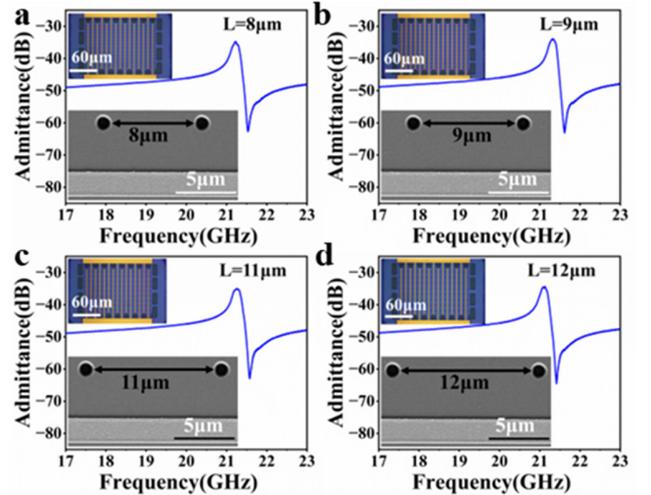

**Fig. 10.** Measured admittance spectra of the A3 resonators with different through-hole distances (a) L=8μm, (b) L=9μm, (c) L=11μm, (d) L=12μm. Insets are (top) optical microscope images of the corresponding resonators and (bottom) zoomed-in SEM images.

As shown in Figs 8-10, the overall trend in the admittance spectra of the resonators remains nearly the same despite variations in through-hole parameters. Table IV further confirms that the key performance parameters of the resonators remain generally consistent, without significant decreases. The optical microscope images demonstrate that the suspension area across all resonators is nearly uniform. This consistency underscores the versatility of the through-hole design, indicating that it does not require highly precise adjustments in design and fabrication.

TABLE IV
EXTRACTED KEY PARAMETERS FROM RESONATORS WITH DIFFERENT THROUGH-HOLE CHARACTERISTICS

| Parameter | $f_s$(GHz) | $k_t^2$ | $Q_s$ | $Q_p$ | $H_{need}$(μm) | Size(μm) |
|---|---|---|---|---|---|---|
| Squares | 21.16 | 3.52% | 106 | 336 | 15 | 160*320 |
| Rhombus | 21.20 | 3.37% | 140 | 335 | 15 | 160*320 |
| Squares-90° | 21.37 | 3.38% | 142 | 373 | 15 | 160*320 |
| Position-offside | 21.3 | 3.42% | 136 | 387 | 15 | 160*320 |
| D=0.50μm | 21.2 | 2.95% | 60 | 166 | 15 | 160*320 |
| D=0.75μm | 21.13 | 3.65% | 89 | 293 | 15 | 160*320 |
| D=1.25μm | 21.21 | 3.54% | 109 | 364 | 15 | 160*320 |
| D=1.50μm | 31.31 | 3.59% | 89 | 296 | 15 | 160*320 |
| L=8μm | 21.26 | 3.50% | 98 | 326 | 13 | 160*320 |
| L=9μm | 21.33 | 3.76% | 100 | 333 | 14 | 160*320 |
| L=11μm | 21.58 | 3.60% | 88 | 342 | 15 | 160*320 |
| L=12μm | 21.21 | 3.43% | 99 | 498 | 16 | 160*320 |

## VI. CONCLUSION

This research introduces and experimentally verifies, for the first time, a Lamb wave resonator featuring sub-wavelength through-holes. Using an LN A3 resonator operating in the K band as an advanced example, it is demonstrated that through-holes can address key challenges faced by Lamb wave resonators without introducing additional processing steps or compromising the resonator's original performance. By introducing sub-wavelength through-holes on the surface of the (LN) piezoelectric film and between the interdigital electrodes, the ineffective suspension area of the Lamb wave resonator will be significantly reduced. This approach is expected to improve the mechanical stability, temperature stability, and power tolerance of both the resonator and the filter. Additionally, the through-hole structure reduces etching time and manufacturing costs. Importantly, it ensures uniform etching distances (*i.e.*, etching times) for resonators with varying structural parameters on a single wafer, which is crucial for the large-scale production of Lamb filters. The LN A3 resonators with through-holes demonstrated in this paper, fabricated using a 260 nm thick Z-cut LN single crystal film, operates at frequencies exceeding 21 GHz, achieves an electromechanical coupling coefficient greater than 4%, quality factor in the hundreds, with no spurious modes. Beyond the resonators discussed here, through-hole structures could enhance Lamb wave resonators using various piezoelectric films, such as LiTaO$_3$ [26][27], AlN [28-30], and AlScN [31][32], as well as other acoustic plate modes [33][34]. This method has the potential to significantly advance the practical implementation of Lamb wave resonators, paving the way for breakthroughs in next-generation wireless filtering, IoT sensing, and other cutting-edge technological domains.

REFERENCES

[1] F. A. Miranda, G. Subramanyam, F. W. van Keuls, R. R. Romanofsky, J. D. Warner and C. H. Mueller, "Design and development of ferroelectric tunable microwave components for Kuand K-band satellite communication systems," *IEEE Trans. Microw. Theory Techn,* vol. 48, no. 7, pp. 1181-1189, July 2000, doi: 10.1109/22.853458.

[2] X. Huang, X. Zhang, L. Zhou, J. -X. Xu and J. -F. Mao, "Low-Loss Self-Packaged Ka-Band LTCC Filter Using Artificial Multimode SIW Resonator," *IEEE Transactions on Circuits and Systems II: Express Briefs*, vol. 70, no. 2, pp. 451-455, Feb. 2023, doi: 10.1109/TCSII.2022.3173712.

[3] G. Basavarajappa and R. R. Mansour, "An Efficient EM-Based Synthesis Technique for Single-Band and Dual-Band Waveguide Filters," *IEEE Transactions on Computer-Aided Design of Integrated Circuits and Systems*, vol. 41, no. 6, pp. 1687-1692, June 2022, doi: 10.1109/TCAD.2021.3093016

[4] J. Zhang, J. -X. Xu, C. Yao and X. Y. Zhang, "Miniaturized High-Selectivity High-Resistivity-Silicon IPD Bandpass Filter Based on Multiple Transmission Paths," *IEEE ELECTR DEVICE L*, vol. 45, no. 4, pp. 534-537, April 2024, doi: 10.1109/LED.2024.3364692

[5] S. Gong, R. Lu, Y. Yang, L. Gao, and A. E. Hassanien, "Microwave Acoustic Devices: Recent Advances and Outlook," *IEEE Journal of Microwaves,* vol. 1, no.2, pp. 601-609,2021, doi: 10.1109/jmw.2021.3064825.

[6] Y. Yang, R. Lu, L. Gao and S. Gong, "10–60-GHz Electromechanical Resonators Using Thin-Film Lithium Niobate," *IEEE Trans. Microw. Theory Technol*, vol. 68, no. 12, pp. 5211-5220, Dec. 2020, doi: 10.1109/TMTT.2020.3027694.

[7] R. Lu and S. Gong, "RF acoustic microsystems based on suspended lithium niobate thin films: advances and outlook," *J MICROMECH MICROENG,* vol. 31, no. 11, 2021, doi: 10.1088/1361-6439/ac288f.

[8] Y. Yang, L. Gao, R. Lu and S. Gong, "Lateral Spurious Mode Suppression in Lithium Niobate A1 Resonators," *IEEE Trans. Ultrason., Ferroelectr., Freq. Control*, vol. 68, no. 5, pp. 1930-1937, May 2021, doi: 10.1109/TUFFC.2020.3049084

[9] Y. Yang, R. Lu, L. Gao and S. Gong, "4.5 GHz Lithium Niobate MEMS Filters With 10% Fractional Bandwidth for 5G Front-Ends," *J. Microelectromech. Sys*, vol. 28, no. 4, pp. 575-577, Aug. 2019, doi: 10.1109/JMEMS.2019.2922935.

[10] Z. Wu, K. Yang, F. Lin and C. Zuo, "6.2 GHz Lithium Niobate MEMS Filter with FBW of 11.8% and IL of 1.7 dB," in *2022 IEEE MTT-S International Conference on Microwave Acoustics and Mechanics (IC-MAM)*, Munich, Germany, 2022, pp. 98-101.

[11] V. Plessky, S. Yandrapalli, P. J. Turner, L. G. Villanueva, J. Koskela, and R. B. Hammond, "5 GHz laterally-excited bulk-wave resonators (XBARs)

based on thin platelets of lithium niobate," *Electron. Lett,* vol. 55, no. 2, pp. 98-100, 2019, doi: 10.1049/el.2018.7297.

[12] S. Yandrapalli, S. E. K. Eroglu, V. Plessky, H. B. Atakan, and L. G. Villanueva, "Study of Thin Film LiNbO$_3$ Laterally Excited Bulk Acoustic Resonators," *J. Microelectromech. Syst,* vol. 31, no. 2, pp. 217-225, 2022, doi: 10.1109/jmems.2022.3143354.

[13] P. J. Turner, B. Garcia, V. Yantchev, G. Dyer, S. Yandrapalli, L. G. Villanueva, R. B. Hammond, and V. Plessky., "5 GHz Band n79 wideband microacoustic filter using thin lithium niobate membrane," *Electron. Lett,* Article vol. 55, no. 17, pp. 942-943, Aug 22 2019, doi: 10.1049/el.2019.1658.

[14] J. Koulakis, J. Koskela, W. Yang, L. Myers, G. Dyer and B. Garcia, "XBAR physics and next generation filter design," *2021 IEEE International Ultrasonics Symposium (IUS)*, Xi'an, China, 2021, pp. 1-5,

[15] R. Lu, Y. Yang, S. Link, and S. Gong, "A1 Resonators in 128° Y-cut Lithium Niobate with Electromechanical Coupling of 46.4%," *J. Microelectromech. Syst,* vol. 29, no. 3, pp. 313-319, 2020, doi: 10.1109/jmems.2020.2982775.

[16] Y. Yang, R. Lu, and S. Gong, "High Q Antisymmetric Mode Lithium Niobate MEMS Resonators With Spurious Mitigation," *J. Microelectromech. Syst,* vol. 29, no. 2, pp. 135-143, 2020, doi: 10.1109/jmems.2020.2967784.

[17] X. Gu *et al.*, "Study of High-Q Laterally Excited Bulk Wave Resonator With Smaller Gap-Width Reflectors," *IEEE ELECTR DEVICE Lett,* vol. 44, no. 8, pp. 1344-1347, 2023, doi: 10.1109/led.2023.3285813.

[18] Y. Yang, L. Gao, and S. Gong, "X-Band Miniature Filters Using Lithium Niobate Acoustic Resonators and Bandwidth Widening Technique," *IEEE Trans. Microw. Theory Technol,* vol. 69, no. 3, pp. 1602-1610, 2021, doi: 10.1109/tmtt.2021.3049434.

[19] F. Lin, K. Yang, and C. Zuo, "A 20.4-GHz Lithium Niobate A3-Mode Resonator with High Electromechanical Coupling of 6.95%," in *2023 IEEE/MTT-S International Microwave Symposium - IMS 2023,* San Diego, CA, USA, 2023, pp. 895-898

[20] L. Gao, Y. Yang, and S. Gong, "Wideband Hybrid Monolithic Lithium Niobate Acoustic Filter in the K-Band," *IEEE Trans. Ultrason., Ferroelectr., Freq. Control*, vol. 68, no. 4, pp. 1408-1417, 2021, doi: 10.1109/tuffc.2020.3035123.

[21] K. Yang, F. Lin, J. Fang, J. Chen, H. Tao, H. Sun, and C. Zuo, "Nanosheet Lithium Niobate Acoustic Resonator for mmWave Frequencies," *IEEE ELECTR DEVICE L,* vol. 45, no. 2, pp. 272-275, 2024, doi: 10.1109/led.2023.3345345.

[22] O. Barrera, S. Cho, L. Matto, J. Kramer, K. Huynh, V. Chulukhadze, Y.-W. Chang, M. S. Goorsky, and R. Lu., "Thin-Film Lithium Niobate Acoustic Filter at 23.5 GHz With 2.38 dB IL and 18.2% FBW," *J. Microelectromech. Syst,* vol. 32, no. 6, pp. 622-625, 2023, doi: 10.1109/jmems.2023.3314666.

[23] J. Xie, M. Shen, Y. Xu, W. Fu, L. Yang, and H. X. Tang, "Sub-terahertz electromechanics," *Nature Electronics*, vol. 6, no. 4, pp. 301-306, 2023, doi: 10.1038/s41928-023-00942-y.

[24] Y. Zhang, Y. Jiang, C. Tang, C. Deng, F. Du, J. He, Q. Hu, Q. Wang, H. Yu, and Z. Wang., "Lithium Niobate MEMS Antisymmetric Lamb Wave Resonators with Support Structures," *Micromachines*, vol. 15, no. 2, 2024, doi: 10.3390/mi15020195.

[25] X. Tong *et al.*, "High figure-of-merit A1-mode lamb wave resonators operating around 6 GHz based on the LiNbO3 thin film," *J PHYS D APPL PHYS,* vol. 57, no. 29, 2024, doi: 10.1088/1361-6463/ad3e06.

[26] Y. Majd and R. Abdolvand, "Designing the Turnover Temperature in Lamb- Wave Lithium Tantalate Resonators," *IEEE ELECTR DEVICE L,* vol. 45, no. 8, pp. 1508-1511, 2024, doi: 10.1109/led.2024.3411403.

[27] N. Assila, M. Kadota, and S. Tanaka, "High-Frequency Resonator Using A1 Lamb Wave Mode in LiTaO3 Plate," *IEEE Trans. Ultrason., Ferroelectr., Freq. Control,* vol. 66, no. 9, pp. 1529-1535, 2019, doi: 10.1109/tuffc.2019.2923579.

[28] J. Zou, A. Gao, and A. P. Pisano, "Ultralow Acoustic Loss Micromachined Butterfly Lamb Wave Resonators on AlN Plates," *IEEE Trans Ultrason Ferroelectr Freq Control,* vol. 67, no. 3, pp. 671-674, Mar 2020, doi: 10.1109/TUFFC.2019.2945235.

[29] G. Chen and M. Rinaldi, "Aluminum Nitride Combined Overtone Resonators for the 5G High Frequency Bands," *J. Microelectromech. Syst,* vol. 29, no. 2, pp. 148-159, 2020, doi: 10.1109/jmems.2020.2975557.

[30] J. Zou, C.-M. Lin, A. Gao, and A. P. Pisano, "The Multi-Mode Resonance in AlN Lamb Wave Resonators," *J. Microelectromech. Syst,* vol. 27, no. 6, pp. 973-984, 2018, doi: 10.1109/jmems.2018.2867813.

[31] S. Shao, Z. Luo, Y. Lu, A. Mazzalai, C. Tosi, and T. Wu, "High Quality Co-Sputtering AlScN Thin Films for Piezoelectric Lamb-Wave Resonators," *J. Microelectromech. Syst,* vol. 31, no. 3, pp. 328-337, 2022, doi: 10.1109/jmems.2022.3161055.

[32] T. Luo, Q. Xu, Z. Wen, Y. Qu, J. Zhou, B. Lin, Y. Cai, Y. Liu, and C. Sun, "Spurious-Free S$_1$ Mode AlN/ScAlN-Based Lamb Wave Resonator With Trapezoidal Electrodes," *IEEE ELECTR DEVICE Lett*, vol. 44, no. 4, pp. 574-577, 2023, doi: 10.1109/led.2023.3244585.

[33] Z. Dai, X. Liu, H. Cheng, S. Xiao, H. Sun, and C. Zuo, "Ultra High Q Lithium Niobate Resonator at 15-Degree Three-Dimensional Euler Angle," *IEEE ELECTR DEVICE Lett*, vol. 43, no. 7, pp. 1105-1108, 2022, doi: 10.1109/led.2022.3175572.

[34] Y. -H. Song and S. Gong, "Wideband RF Filters Using Medium-Scale Integration of Lithium Niobate Laterally Vibrating Resonators," *IEEE ELECTR DEVICE Lett*, vol. 38, no. 3, pp. 387-390, March 2017, doi: 10.1109/LED.2017.2662066